\documentclass[final,1p,times]{elsarticle} 
\usepackage{graphicx} 
\usepackage{amssymb} 
\usepackage{amsthm} 
\usepackage{lineno} 

\journal{Nuclear Physics A} 

\usepackage{epsfig}
\usepackage{amsmath}
\usepackage{graphicx}
\usepackage{bm}

\graphicspath{{Figs/}}

\newcommand{\be}{\begin{equation}}
\newcommand{\ee}{\end{equation}}
\newcommand{\bea}{\begin{eqnarray}}
\newcommand{\eea}{\end{eqnarray}}
\newcommand{\bem}{\begin{multline}}
\newcommand{\eem}{\end{multline}}
\newcommand{\beg}{\begin{gather}}
\newcommand{\eeg}{\end{gather}}

\def\eq#1{{Eq.~(\ref{#1})}}
\def\fig#1{{Fig.~\ref{#1}}}
\newcommand{\ben}{\begin{eqnarray*}}
\newcommand{\een}{\end{eqnarray*}}

\begin{document} 

\begin{frontmatter} 


\title{Heavy Quark Potential at Finite Temperature in AdS/CFT}

\author{Javier L. Albacete$^{a,b}$}

\address[a]{IPhT-Sacaly, Orme des Merisiers batiment 774, F-9119,1 Gif-sur-Yvette Cedex, France.}
\address[b]{ECT*, Strada delle Tabarelle 286, I-38050, Villazzano (TN), Italy.}

\begin{abstract} 
A calculation of the heavy quark potential at finite temperature at strong coupling based on the 
AdS/CFT correspondence is presented. The calculation relies on the method of complex string trajectories 
previously introduced in \cite{Albacete:2008ze}, and on the introduction of  a modified renormalization 
subtraction. The obtained potential is smooth,  negative definite for all quark-antiquark separations, and 
develops an imaginary part for  $r > r_c \!=\! 0.870/\pi T$ . At large separations the real part of the
  potential does not exhibit the exponential Debye falloff expected
  from perturbation theory and instead falls off as a power law,
  proportional to $1/r^4$.
\end{abstract} 

\end{frontmatter} 

\linenumbers 

The heavy quark potential at finite temperature is a very important quantity in the study of Quark Gluon 
Plasma formation in Heavy Ion Collisions carried out at RHIC. Until recently it has been calculated either
analytically at small coupling using perturbation theory, or
numerically using lattice simulations, hence in Euclidean time. However, the matter formed in RHIC collisions is 
rapidly evolving in time and, according to latest analyses, it is strongly coupled. Thus, calculational tools for 
strongly coupled, real-time QCD dynamics are needed. Such tools are now available via the Anti-de-Sitter 
space/conformal field theory (AdS/CFT) correspondence, albeit for a theory different to QCD, namely $
\mathcal{N}=4$ supersymmetric Yang-Mills theory. 
 
The first calculation of a heavy quark potential in vacuum for ${\cal
  N} =4$ SYM theory was carried out by Maldacena in
\cite{Maldacena:1998im}. There, the potential is obtained from
the expectation value of a static temporal Wilson loop which, on the gravity side, corresponds to the action of 
the worldsheet spanned by an open string connecting the quark and the antiquark. Due to conformal invariance of $
{\cal N} =4$ SYM theory, the potential is of Coulomb type:
\begin{align}\label{zeroT}
  V_0 (r) \, = \, - \frac{\sqrt{\lambda}}{2 \, \pi \, c_0^2 \, r}
\end{align}
with $r$ is the distance
between the quark and the anti-quark, $\lambda$ the 't Hooft coupling and $c_0 \, = \, \Gamma^2
\left(\frac{1}{4}\right)/(2 \, \pi)^{3/2}$. Soon after \cite{Maldacena:1998im}
calculations of the heavy quark potential for ${\cal N} =4$ SYM theory
at finite temperature appeared in \cite{Rey:1998bq,Brandhuber:1998bs}.
The potential obtained in
these calculations  starts out at small $r$ being
close to the vacuum potential of \eq{zeroT}, but rises steeper than
the vacuum potential, becoming zero at a separation $r^* = 0.754 /\pi
T$. For larger separations, i.e., for $r>r^*$, the authors of
\cite{Rey:1998bq,Brandhuber:1998bs} argue that the string ``melts'',
and the dominant configuration corresponds to two straight strings
stretching from the quark and the anti-quark down to the black hole
horizon. The resulting potential is thus zero for $r>r^*$ and has a
kink (a discontinuity in its derivative) at $r = r^*$. Here I summarize the results presented in \cite{Albacete:2008dz}, 
where some modifications to the calculations in \cite{Rey:1998bq,Brandhuber:1998bs} were proposed in order 
to avoid the unphysical features of the resulting finite-$T$ potential, namely its trivial infrared behavior plus the presence 
of discontinuities in its derivative (which leads to infinite forces). 

In $SU(N_c)$ gauge theory in Euclidean time, the heavy quark potential or, more 
precisely, the free energy, is calculated through the connected correlator of two Polyakov 
loops at spatial separation $\vec{r}$:
\begin{align}\label{L}
  \langle L (0) \, L^\dagger ({\vec r}) \rangle_c = \frac{e^{- \beta
      \, V_1 (r) } + (N_c^2 -1) \, e^{- \beta \, V_{adj} (r)
    }}{N_c^2}\,\,\, ; \,\,\, L ({\vec r})  =  \frac{1}{N_c} \, \text{Tr} \left[ \text{P} \exp
    \left( i g \!\!\int_0^\beta d \tau \, A_4 ({\vec r}, \tau )\right)
  \right],
\end{align}
where $\beta=1/T$. \eq{L} is the definition of singlet $V_1 (r)$ and adjoint $V_{adj}
(r)$ potentials. In the AdS/CFT set up, the color decomposition of \eq{L}  can be read off from the different 
open string configurations attached to the quark and the antiquark. One can have a "hanging" string connecting 
the quark and the antiquark or, alternatively, two "straight" strings going from the (anti)quark to 
the black hole horizon. $N_c$-counting in AdS indicates that the hanging string configuration
gives $V_1 (r)$, since there is just one way to connect the quark and the antiquark, while the two straight 
strings give $V_{adj} (r)$, since there are $N_c^2-1$ different ways in which two independent strings can be 
connected to the $N_c$ D3-branes living at the bottom of the AdS space. Moreover, the two straight strings 
can only be connected through graviton exchange in the bulk. Therefore the contribution of the two straight string configuration to the potential is $1/N_c^2$ suppressed, in agreement with the analysis of \cite{Nadkarni:1986cz}

We calculate the singlet
potential $V_1 (r)$ according to the definition proposed in \cite{Laine:2006ns} in the real-time formalism: $V_1 
(r)$ is given by the expectation value of a static (temporal) Wilson loop via
\begin{align}
  \langle W \rangle \, = \, e^{- i \, {\cal T} \, V_1 (r)}
\end{align}
with the temporal extent of the Wilson loop $\cal T \rightarrow
\infty$. To calculate the Wilson loop in the AdS/CFT set up, we start with the AdS$_5$ black
hole metric in Minkowski space
\begin{align}
  ds^2 = \frac{L^2}{z^2} \left[ - \left( 1 - \frac{z^4}{z_h^4} \right)
    dt^2 + d {\vec x}^2 + \frac{d z^2}{1 - \frac{z^4}{z_h^4}} \right]\,,
\end{align}
where $d {\vec x}^2 = (d x^1)^2 + (d x^2)^2 +(d x^3)^2$, $z$ is the
coordinate describing the 5th dimension and $L$ is the curvature of
the AdS$_5$ space. The horizon of the black hole is located at $z =
z_h$ with $z_h = 1/\pi T$. We want to extremize the open string worldsheet for a string attached
to a static quark at $x^1 = r/2, x^2 = x^3 =0$ and an anti-quark at
$x^1 = -r/2, x^2 = x^3 =0$. Parameterizing the static string
coordinates by  $X^\mu \! = \! \left[ X^0=t, X^1=x , X^2=0, X^3=0, X^4=z(x) \right]$ we write the Nambu-Goto 
(NG) action as 
\begin{align}\label{NG1}
  S_{NG} (r, T) \, = \, - \frac{\sqrt{\lambda}}{2 \, \pi} \, {\cal T}
  \int_{-r/2}^{r/2} dx \, \sqrt{\frac{1+z'^2}{z^4} - \frac{1}{z_h^4}},
\end{align}
where $z' = d z(x) /dx$. Finally, the action in \eq{NG1} contains a UV divergence associated to the infinite 
mass of the quarks which has to be
subtracted out. Usually, the subtraction contains a finite piece as
well \cite{Maldacena:1998im,Rey:1998bq,Brandhuber:1998bs}, which may
be temperature-dependent in the case at hand. Here we will use the
following subtraction:  We define the quark--anti-quark potential by 
\begin{align}\label{subtr}
  V (r) \, = \, - \left\{ S_{NG} (r, T) - \text{Re} [S_{NG} (r =
    \infty, T)] \right\} / {\cal T}.
\end{align}
Our prescription is different from the one used in
\cite{Rey:1998bq,Brandhuber:1998bs}, where the contribution of two straight strings were subtracted. It 
insures that the real part of the potential $V(r)$
goes to zero at infinite separations. The definition in \eq{subtr} is
consistent with the original prescription outlined in \cite{Maldacena:1998im} to find the heavy
quark potential at zero temperature. Importantly, it is also consistent with our choice of color 
representation, since only {\it singlet} configurations appear in the definiiton of $V(r)$.

What follows next is just a problem of classical mechanics: One has to find the string 
trajectories that minimize 
the NG action \eq{NG1}. In order to do so one has to derive and solve the Euler-Lagrange 
equations with the 
appropriate boundary conditions,  $z(x=\pm r/2)=0$. The details of this calculation are 
explained in \cite
{Albacete:2008dz}. Importantly, the classical solutions can be parametrized in terms of the 
quark--
antiquark separation $r$, the temperature $T$ (both "external" parameters), and the 
maximum of the string 
along the 5th dimension, $z_{max}$. The later is determined by the condition $z'_{max}
(r,x=0)=0$.  We obtain the following expression for the heavy quark potential of the $
{\cal N} =4$ SYM theory at finite temperature:
\begin{align}\label{pot}
  V (r) \, = \, \frac{\sqrt{\lambda}}{2 \, c_0 \, \pi} \, \bigg[ - &
  \frac{1}{z_{max}} \, \left( 1 - \frac{z_{max}^4}{z_h^4} \right)
   F \left( \frac{1}{2}, \frac{3}{4} ;
    \frac{1}{4} ; \frac{z_{max}^4}{z_h^4} \right) + \frac{1}{z_h}
  \bigg],
\end{align} 
where $F$ is the hypergeometric function. Importantly, the solutions for $z_{max}$
become complex for $r > r_c = 0.870 \, z_h$, leading to
complex-valued string trajectories $z(x)$ and potential $V(r)$. This led the authors
of \cite{Rey:1998bq,Brandhuber:1998bs} to abandon their solution for
$r > r_c$. We suggest however to interpret the complex-valued saddle
points as corresponding to quasi-classical configurations in the
classically forbidden region of string coordinates, as previously suggested in \cite{Albacete:2008ze}. This is 
similar to the method of complex trajectories used in quasi-classical
approximations to quantum mechanics.

\begin{figure}[h]
  \centering 
\includegraphics[width=6.cm]{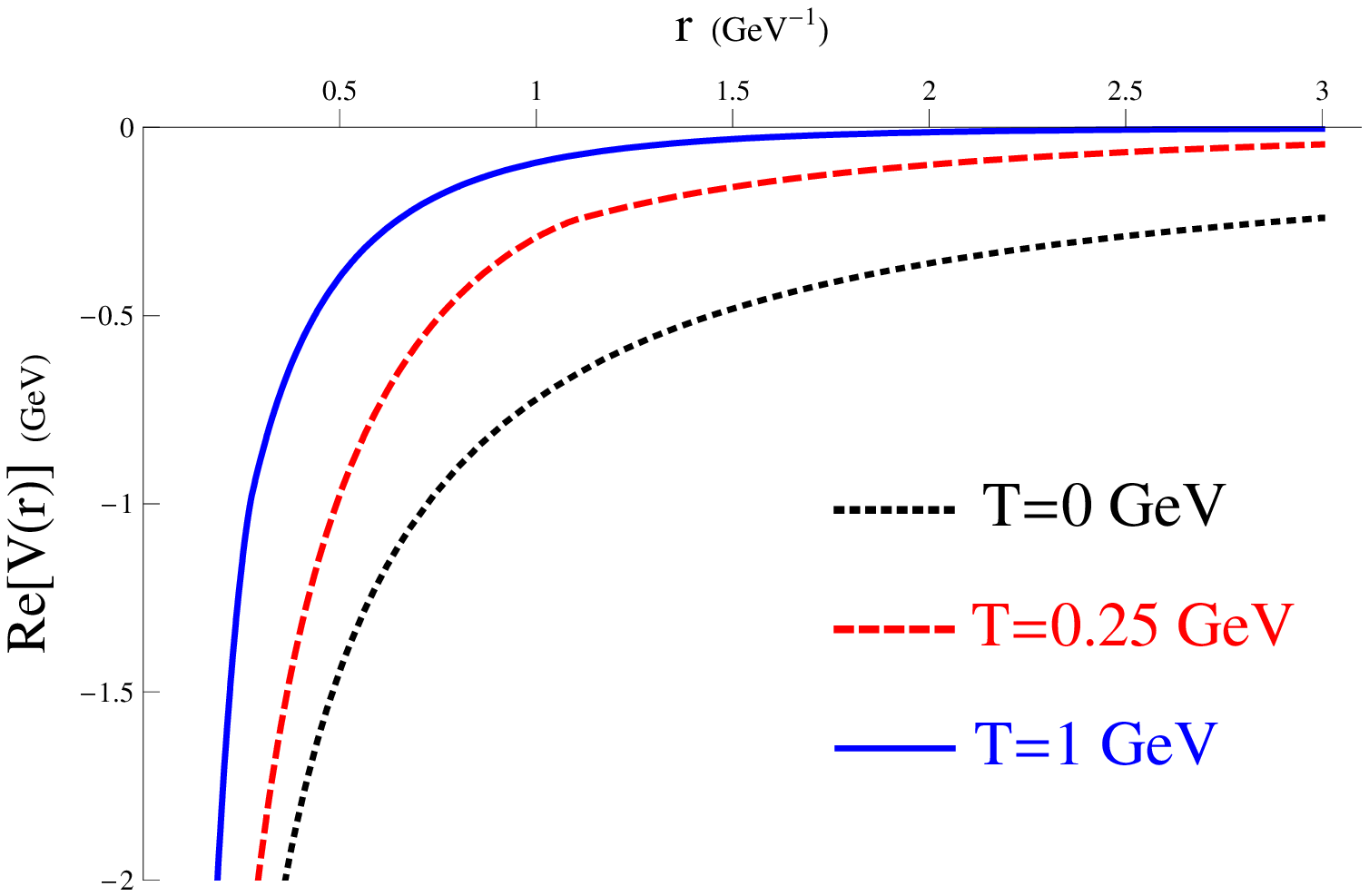}
\includegraphics[width=6.cm]{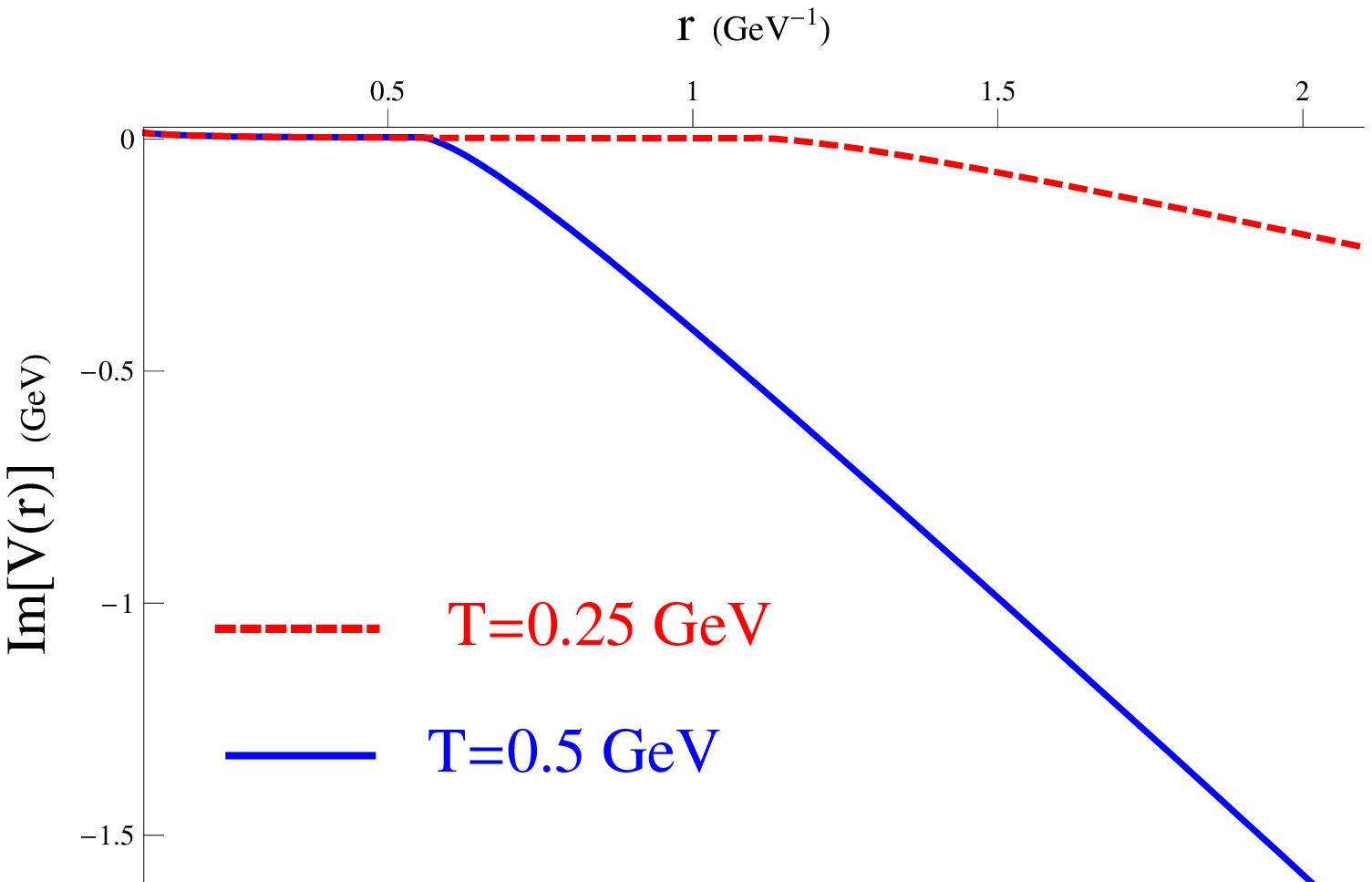}
\caption{The real and imaginary parts of the heavy quark potential 
  plotted as functions of the separation $r$ for several different
  temperatures. We put $\lambda =10$.~\vspace*{-7.8mm}}
\label{fig2}
\end{figure}

Moreover, the solution for  $z_{max}$ has several branches.  In order to select the physical branch we impose 
two conditions: $i)$ The right branch should map onto Maldacena's solution at small $r$. $ii)$ In order to get a 
sensible the quantum-mechanical time-evolution we require that $\text{Im} [V(r)] < 0$, sot that the probability 
of a state, $\sim e^{\text{Im} [E]  \, t}$  does not exceed unity. These criteria suffice to single just one branch 
for $z_{max}$. Inserting it into \eq{pot}, we readily obtain the real and imaginary parts of the resulting potential, plotted in \fig{fig2} for two non-zero temperatures, along
with the zero-T curve for comparison.

\vskip 0.7 cm
Several comments are in order: First, the real part of the potential is a smooth, negative definite function, going to zero at large inter-quark separations. This latest feature is a direct consequence of our renormalization prescription \eq{subtr}. If we had used the renormalization proposed in \cite{Rey:1998bq,Brandhuber:1998bs}, then $\text{Re} [V(r)]$ would have gone to a positive constant as $r \rightarrow \infty$.
Moreover, the non-zero
temperature curves exhibit a strong screening compared to the zero
temperature case, while at small $r$ we recover the zero temperature potential of \cite{Maldacena:1998im}. 
On the other hand, the heavy quark potential develops an imaginary part for $r > r_c$. This 
means the potential becomes absorptive, as the $q\bar q$ singlet state may decay in the medium, with an 
absorption rate that increases with the quark--antiquark separation. The existence of an imaginary part in
the heavy quark potential has been previously observed in perturbation
theory in \cite{Laine:2006ns}.

The nature of the screening exhibited by the lines in \fig{fig2} can be better understood 
from the following asymptotic expansions for the real and imaginary part of the potential:
\begin{align}\label{rvexp}
  \text{Re} [V (r)]\bigg|_{r \, T \gg 1} \, = \, - \frac{\pi^3 \,
    c_0^3}{4} \, \sqrt{\lambda} \, \frac{z_h^3}{r^4} + o \left(
    \frac{z_h^4}{r^5} \right)\,,\,\,\text{and}\,\, \text{Im} [V (r)] \bigg|_{r \, T \gg 1} \! \! =  -
  \frac{\sqrt{\lambda}}{\pi} \, \frac{1}{2 \, z_h} \, \left[
    \frac{r}{z_h} - \frac{1}{c_0} + o \left( \frac{z_h}{r} \right)
  \right]. 
\end{align}
Instead of the exponential falloff with $r$ characteristic of Debye screening expected from 
perturbation theory and which has been postulated for ${\cal N} =4$ SYM theory at strong 
coupling in
\cite{Bak:2007fk}, the real part of the heavy quark potential falls
off as a power, $\text{Re} [V (r)] \sim 1/T^3 r^4$, at large $r$. If our hypothesis of using 
the complex string configurations is confirmed, this would be an interesting new type of 
screening for the potential. However, the large negative imaginary part of the potential 
leads to exponential decay with time of the heavy quark pair, which may blur the potential 
observable consequences of such new screening mechanism. 
Also, combining the large- and small-$r$ asymptotics we can interpolate the real part of 
the potential to
write an approximate formula
\begin{align}\label{approx}
  \text{Re} [V (r)] \approx - \frac{\sqrt{\lambda}}{2 \, \pi \, c_0^2
    \, r} \, \frac{r_0^3}{(r_0 + r)^3},\quad\text{with}\quad r_0 \, = \, z_h \, \pi \, c_0 \, \left
    ( \frac{\pi \, c_0^2}{2}
  \right)^{1/3} \, \approx \, \frac{2.702}{\pi \, T},
\end{align}
where the parameter $r_0$  given can be interpreted as the screening length.

Several concluding remarks: First, Debye screening is the result 
of a perturbative calculation. Therefore, it may not be valid for distances of the order of $r\sim 1/(g^2\,T)$, where non-perturbative effects may be important. On 
the other hand, one should also bear in mind that the particle content of $\mathcal{N}=4$ SYM is quite 
different to that of QCD. In particular it does not include particles in the fundamental representation (i.e. 
quarks), but only in the adjoint (The heavy quarks in our calculation ought to be regarded as external test 
charges). Since a charge in the adjoint cannot fully screen a fundamental charge, it is reasonable to expect a 
milder screening. Finally, the power-law fall off of the heavy quark potential has also been suggested in 
perturbative calculations \cite{Gale:1987en}. 



\end{document}